\documentclass[12pt, preprint]{aastex}
\usepackage{txfonts}
\usepackage{natbib}
\usepackage{graphicx}
\usepackage{rotating}

\slugcomment{Not to appear in Nonlearned J., 45.}

\newcommand{\Ha}{H$\alpha$}

\newcommand{\km}{kms$^{-1}$}

\begin{document}


\title{Spatially Resolved Observations of the Bipolar Optical Outflow from the Brown Dwarf 2MASSJ12073347-3932540. \altaffilmark{1}}


\author{E.T. Whelan\altaffilmark{2,3}}

\author{T.P. Ray\altaffilmark{2}}

\author{F. Comeron\altaffilmark{4}}

\author{F. Bacciotti\altaffilmark{5}}

\author{P.J.  Kavanagh\altaffilmark{6}}

\altaffiltext{1}{Based on data collected by UVES (281.C-5025(A)) and FORS1 (380.C-0412(A)) observations at the VLT on Cerro Paranal (Chile) which is operated by the European Southern Observatory (ESO). Data collected from the ESO archive (071.C-0432(A), 085.C0238(A), 078.C-0158(A, B)) is also used in the analysis presented here. }
\altaffiltext{2}{Dublin Institute for Advanced Studies, School of Cosmic Physics, 31 Fitzwilliam Place, Dublin 2, Ireland}
\altaffiltext{3}{UJF-Grenoble 1 / CNRS-INSU, Institut de Plane?tologie et dÕAstrophysique de Grenoble (IPAG) UMR 5274, Grenoble, F-38041,
France}
\altaffiltext{4}{ESO, Karl-Schwarzschild-Strasse 2, 85748 Garching, Germany}
\altaffiltext{5}{INAF - Osservatorio Astrofisico di Arcetri, Largo E. Fermi 5, 50125 Firenze, Italy}
\altaffiltext{6}{Institut f\"{u}r Astronomie und Astrophysik, Kepler Center for Astro and Particle Physics, Eberhard Karls Universit\"{a}t, 72076 T\"{u}bingen, Germany}


\begin{abstract}

Studies of brown dwarf (BD) outflows provide information pertinent to questions on BD formation, as well as allowing outflow mechanisms to be investigated at the lowest masses. Here new observations of the bipolar outflow from the 24 M$_{JUP}$ BD, 2MASSJ12073347-3932540 are presented. The outflow was originally identified through the spectro-astrometric analysis of the [OI]$\lambda$6300 emission line. Follow-up observations consisting of spectra and [SII], R-band and I-band images were obtained. The new spectra confirm the original results and are used to constrain the outflow PA at $\sim$ 65$^{\circ}$. The [OI]$\lambda$6300 emission line region is spatially resolved and the outflow is detected in the [SII] images.  The detection is firstly in the form of an elongation of the point spread function along the direction of the outflow PA.  Four faint knot-like features (labelled {\it A-D}) are also observed to the south-west of 2MASSJ12073347-3932540 along the same PA suggested by the spectra and the elongation in the PSF. Interestingly, {\it D}, the feature furthest from the source is bow-shaped with the apex pointing away from 2MASSJ12073347-3932540. A color-color analysis allows us to conclude that at least feature {\it D} is part of the outflow under investigation while {\it A} is likely a star or galaxy. Follow-up observations are needed to confirm the origin of {\it B} and {\it C}. This is a first for a BD, as BD optical outflows have to date only been detected using spectro-astrometry. This result also demonstrates for the first time that BD outflows can be collimated and episodic.

\end{abstract}


\section{Introduction} 
The brown dwarf (BD) 2MASSJ12073347-3932540 (hereafter 2M1207A), with an estimated mass of 24~M$_{JUP}$ \citep{Mohanty07}, has garnered much attention in recent years. Not only does it exhibit young star-like properties, i.e. it is a strong accretor with a detected accretion disk  \citep{Riaz12, Riaz12err, Riaz08, Mohanty03, Mohanty05} and bipolar outflow \citep{Whelan07}, but it also has a planetary mass companion 2M1207B.   2M1207B was first detected by \cite{Chauvin04} using direct imaging and it is one of only a small number of exoplanets to have been detected in this way. The 2M1207 system is a member of the TW Hya association (TWA). TWA is one of the closest young stellar associations to the Sun with an estimated distance of 47-67~pc and an age of $\sim$ 10~Myr \citep{Barrodo06}. Thus TWA is a key reference for the study of star, BD and planet formation and associated phenomena. It is the interesting properties of 2M1207 combined with its mass at the lower end of the BD mass spectrum and relatively close proximity, that have made this object the target of studies aimed at understanding BD and planet formation. 

In this paper new observations of the bipolar outflow driven by 2M1207A are presented. As studies of outflow properties can provide information on the central engine of the driving source, observations of BD outflows can contribute much to the debate on the origin and evolution of BDs.  For example, the kinematics and position angle (PA) of the outflow can be used to infer the disk inclination and PA \citep{Whelan09b}. Episodic or knotty outflows point to variable accretion activity and precession or wiggling in an outflow can be interpreted as evidence of an unresolved companion \citep{Whelan10}. In addition, measurements of the mass outflow to accretion rate ($\dot{M}_{out}$/$\dot{M}_{acc}$) are relevant to jet launching models which in turn affect models of disk evolution and thus planet formation \citep{Ferreira2006, Hartigan1995}. One possible difference between BDs and low mass stars is that Òfirst resultsÓ indicate that $\dot{M}_{out}$/$\dot{M}_{acc}$ may be higher for sub-stellar objects than low mass stars. For low mass stars it is currently estimated at $\sim$ 10$\%$, however, in BDs this ratio has been found to be significantly higher and, in the most extreme cases, the two rates are found to be comparable \citep{Bacciotti2011, Whelan09b}. This is significant as it may help to explain why BDs never accrete enough matter to become stars \citep{Machida09}. Better observations of a larger number of BDs are needed to confirm these results, to properly understand BD accretion/outflow properties and thus to compare BD properties with low mass stars.

BD outflows have mainly been studied through spectroscopy and especially the spectro-astrometric analysis of their forbidden emission line (FEL) regions. The first indications that BDs could drive outflows came from the detection of FELs in their optical spectra \citep{Fernandez01}. Spectro-astrometry (SA), which involves using Gaussian fitting to measure the centroid of an emission region as a function of wavelength/velocity, was used to confirm that these FEL regions were extended and originated in an outflow \citep{Whelan05}. Since the first detection, four further young BD candidates have been found to drive outflows using this method. Imaging studies of BD outflows have rarely been conducted and the mapping of the molecular outflow from the BD ISO-Oph 102 with the sub-millimeter array (SMA), is the highest quality image in existence at present. The very low mass star (VLMS) Par-Lup3-4 with a mass just above the substellar limit \citep{Comeron03} has been studied with a combination of imaging and spectroscopy, and a bipolar outflow was detected using both methods \citep{Comeron11, Fernandez05}. Previous to this work it was the lowest mass protostar where an optical outflow had even been marginally detected using imaging. For Par-Lup3-4 the [SII] image is well extended along the outflow PA and a Herbig Haro (HH) object (HH~600) is also detected. \cite{Comeron11} estimate a proper motion of 168 $\pm$ 30~kms$^{-1}$ for HH~600. \cite{Wang06} carried out a survey for HH objects in the Chamaeleon I molecular cloud with the aim of searching for optical outflows from BDs. While 18 HH objects and $\sim$ 40 BD candidates were identified, they found no persuasive evidence that any of the BDs were related to the HH objects. 

Optical spectra of 2M1207A obtained in May 2006 contained double peaked [OI]$\lambda$6300 emission which was tracing a bipolar outflow \citep{Whelan07}. This was an exciting discovery as it confirmed 2M1207A as the lowest mass galactic object known to drive an outflow. 
The observations presented here can be divided into three separate results. Firstly, new optical spectra confirm the original detection and a comparison with the 2006 spectra allows an estimate of the outflow PA to be made. Secondly, we present a [SII] image in which the outflow is detected close to the source in the form of an elongation in the point spread function (PSF). Thirdly, a series of knot-like features are seen in the [SII] image along the PA suggested by the spectro-astrometric study. We explore the possibility that these features are shocks in the 2M1207A outflow. Below these new observations are described, their significance discussed and the best direction for future studies of BD outflow activity considered.

\section{Observations and Data Analysis}
Both spectroscopic and imaging studies of the outflow driven by 2M1207A have been conducted with the European Southern Observatory's (ESO) Very Large Telescope (VLT), using the UV-Visual Echelle Spectrometer (UVES) and the FOcal Reducer/low-dispersion Spectrograph 1 (FORS-1) \citep{Dekker00, App00, Szei98}. 
Follow-up observations to the original UVES data were made with FORS1 in January/March 2008 and UVES in August 2008. Starting first with the UVES 2008 spectroscopy, observations were carried out with the CD3 grating and a central wavelength of 6000~\AA, providing spectra with a wavelength range of 4900-7100 \AA\ and spectral resolution R=40,000. The slit width and pixel scale were 1\arcsec\ and 0\farcs182, respectively, and the seeing was $\sim$ 0\farcs8. Two observations with exposure time totaling 2900~s were taken. The UVES 2006 data had the same instrumental set-up except that the central wavelength was set at 5800~\AA\ meaning that the wavelength range included the H$\beta$ line at 4861~\AA. Indeed, 2M1207A has been observed numerous times with UVES and the spectra are available in the ESO data archive. However, observations were generally aimed at studying accretion and were not optimised for outflow studies. This means that the exposure time was too low to detect the forbidden emission. 

\begin{figure*}
   \includegraphics[width=15cm]{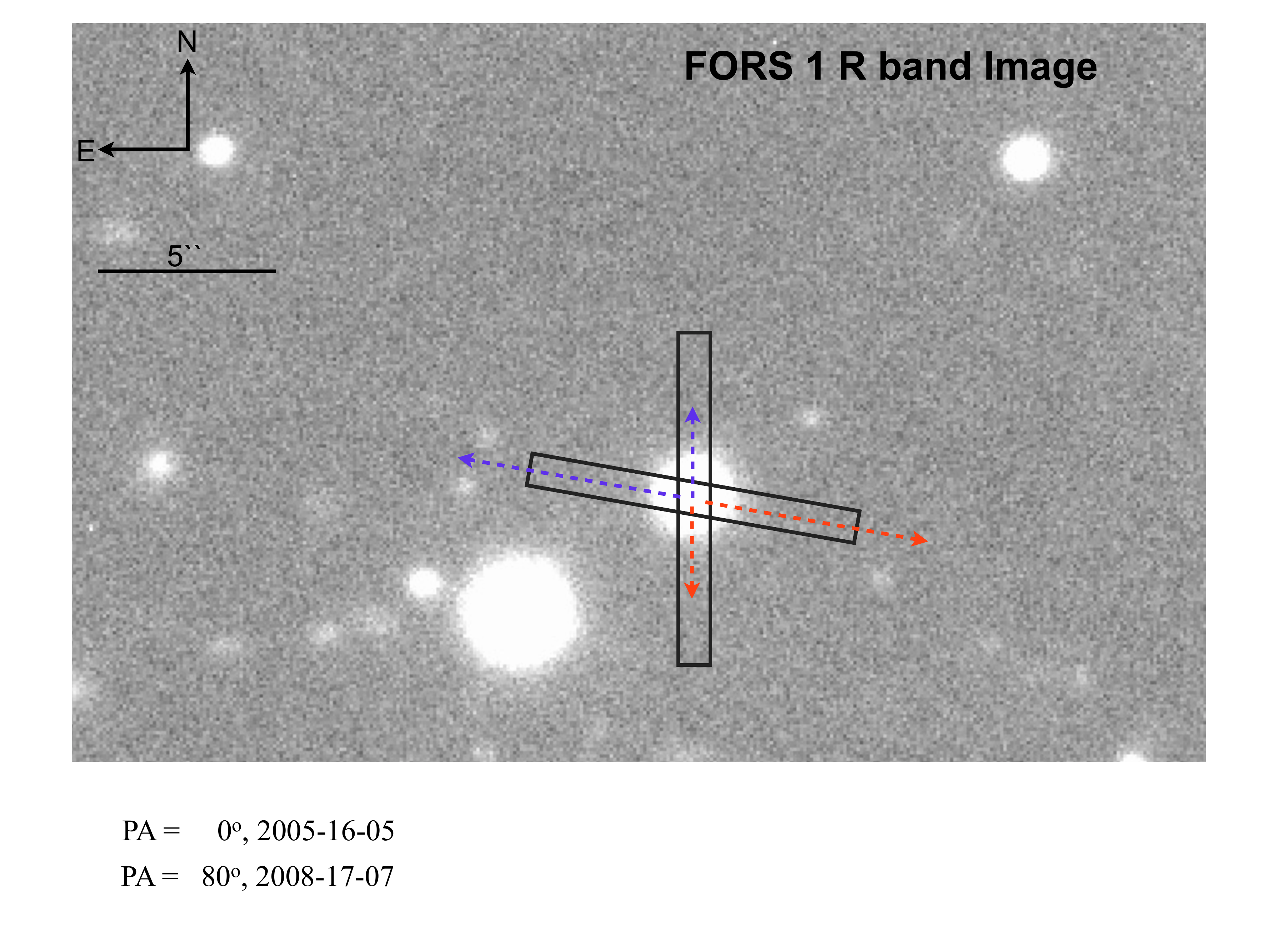}
     \caption{FORS1 R-band image of 2M1207 with the UVES slits and slit PAs shown. The slits are drawn to scale. The spectro-astrometric offsets measured at the two slit PAs, for the blue and red-shifted [OI]$\lambda$6300 emission, are represented by the blue and red arrows. The size of the arrows do not represent the magnitude of the measured offsets. However, they accurately represent the relative magnitudes between the two observations, in that the offsets measured at PA = 80$^{\circ}$ are approximately twice those measured at PA = 0$^{\circ}$.}
     \label{pas}
\end{figure*}

The UVES 2006 observations, published in \cite{Whelan07}, were made with the slit aligned with a PA of 0$^{\circ}$, i.e N-S. For the 2008 spectra the PA was set at the parallactic angle which varied between 75$^{\circ}$ and 85$^{\circ}$ over the course of the observation. By comparing the analysis of the 2008 spectra with the 2006 spectra the PA of the outflow can be probed and, for the purpose of this we take the slit PA to be 80$^{\circ}$. Figure \ref{pas} presents the different slit PAs used and the direction on the sky to which the offsets correspond to. The spectra were reduced using standard routines for bias subtraction, flat fielding and wavelength calibration, provided by the image reduction and analysis facility (IRAF). In order to analyse the spectra using SA, the 2D spectra corresponding to each individual order had to be extracted and wavelength calibrated using the {\it Identify}, {\it Reidentify}, {\it Fitcoords} and {\it Transform} routines. Continuum subtraction was carried out using the {\it Continuum} routine. Overall the process followed for the SA was the same as outlined in \cite{Whelan07} including the smoothing of the data using {\it Gauss} to increase the S/N ratio. Also see \cite{Whelan08} for further details on the spectro-astrometric technique.

Imaging with FORS1 was carried out using the [SII]+62 narrowband filter and the R-BESS and I-BESS broadband filters in March 2008. All observations were made with the high resolution (HR) collimator and a binning of 2$\times$2,  giving a pixel scale of 0\farcs125. Again standard IRAF routines were used for bias subtraction and flat-fielding. For the [SII] images a total of 2 $\times$ 30~mins exposures were made. The exposures were taken consecutively, however the seeing deteriorated from $\sim$ 0\farcs6 to 0\farcs9 between the two exposures. The total exposure time for the broadband images was 15~mins. Although the [OI]$\lambda$6300 line had previously been detected in the 2M1207A spectrum and shown to trace an outflow, for the purposes of imaging, [SII] was preferred over [OI] due to the relatively bright background emission in [OI] from the atmosphere. To help with the analysis of the [SII] FORS1 images, SUSI2 I-band images were obtained from the ESO archive and are further discussed in the Section 3.

As dithering was not used to obtain the I-band images, fringing effects which could not be removed using the flat fields presented a problem to the analysis of faint emission in the vicinity of 2M1207A. To remove the fringing, separate I-band observations were identified and downloaded from the ESO archive. The closest ones in time were taken on 
11 February 2008, which is about one month before our observations of 
2M1207, and had been originally obtained to image SN 2007Y in the galaxy 
NGC 1187. The galaxy has a relatively large angular size that fills a 
significant fraction of the field of view of FORS1. However, the area of the 
detector where 2M1207 was later imaged corresponded to a region in the 
distant outskirts of the galaxy with no obvious structure. Thanks to 
this, it was possible to construct a fringe pattern frame suitable for 
the region immediately surrounding 2M1207 by combining and 
median-filtering the eight available exposures of NGC 1187, uncorrected 
for telescope offsets between the exposures. The fringe pattern 
obtained was visually confirmed to be very similar in appearance to that 
in the 2M1207 exposures. Next, an average sky background value near the 
position of 2M1207 in the detector was subtracted from both the 2M1207 
and fringing pattern images. The background-subtracted fringing pattern 
image was then scaled by a numerical factor and subtracted from the 
background-subtracted image of 2M1207. Based on visual inspection of the 
resulting image, this last part of the process was iterated with 
different scaling factors applied to the background-subtracted fringing 
image, minimizing the residual fringing pattern in the image of 
2M1207. As dithering was used to obtain the SUSI2 images of 2M1207 
identified in the ESO archive, construction of a fringing pattern from 
separate images was not needed in this case. Instead, the frames 
containing the target were combined with median filtering to construct a 
sky frame, including fringing, to be subtracted from the individual 
images. The sky background remained very stable during all the SUSI2 
observations, thus leaving no noticeable fringing residuals in the 
sky-subtracted images.

\section{Results}

\subsection{Spectroscopy and Spectro-astrometry}
\begin{figure*}
   \includegraphics[width=18cm]{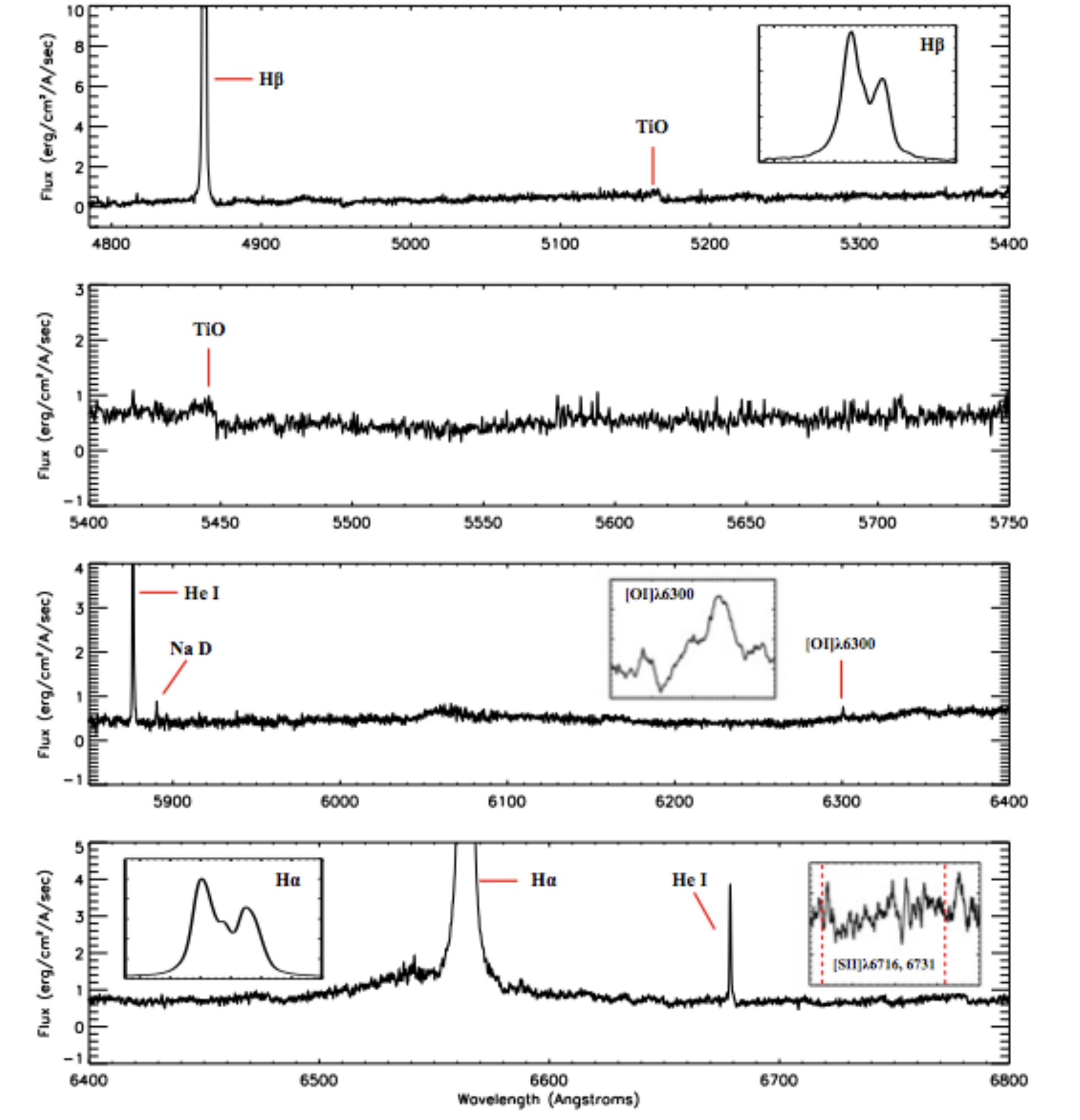}
     \caption{The full spectral range of the UVES observations aimed at studying the 2M1207A outflow. This 1D spectrum was extracted from the source position using the ESO UVES pipeline and was taken on May 16th 2006. H$\alpha$ and H$\beta$ are by far the strongest lines and while [OI]$\lambda$6300 line is detected no [SII] emission is detected above the noise at the source position. In the inset showing the region of the [SII]$\lambda\lambda$6716, 6731 lines, the dashed red lines mark the 0 \km\ position for the two emission regions. }
     \label{spec1d}
\end{figure*}

\begin{figure*}
   \includegraphics[width=18cm]{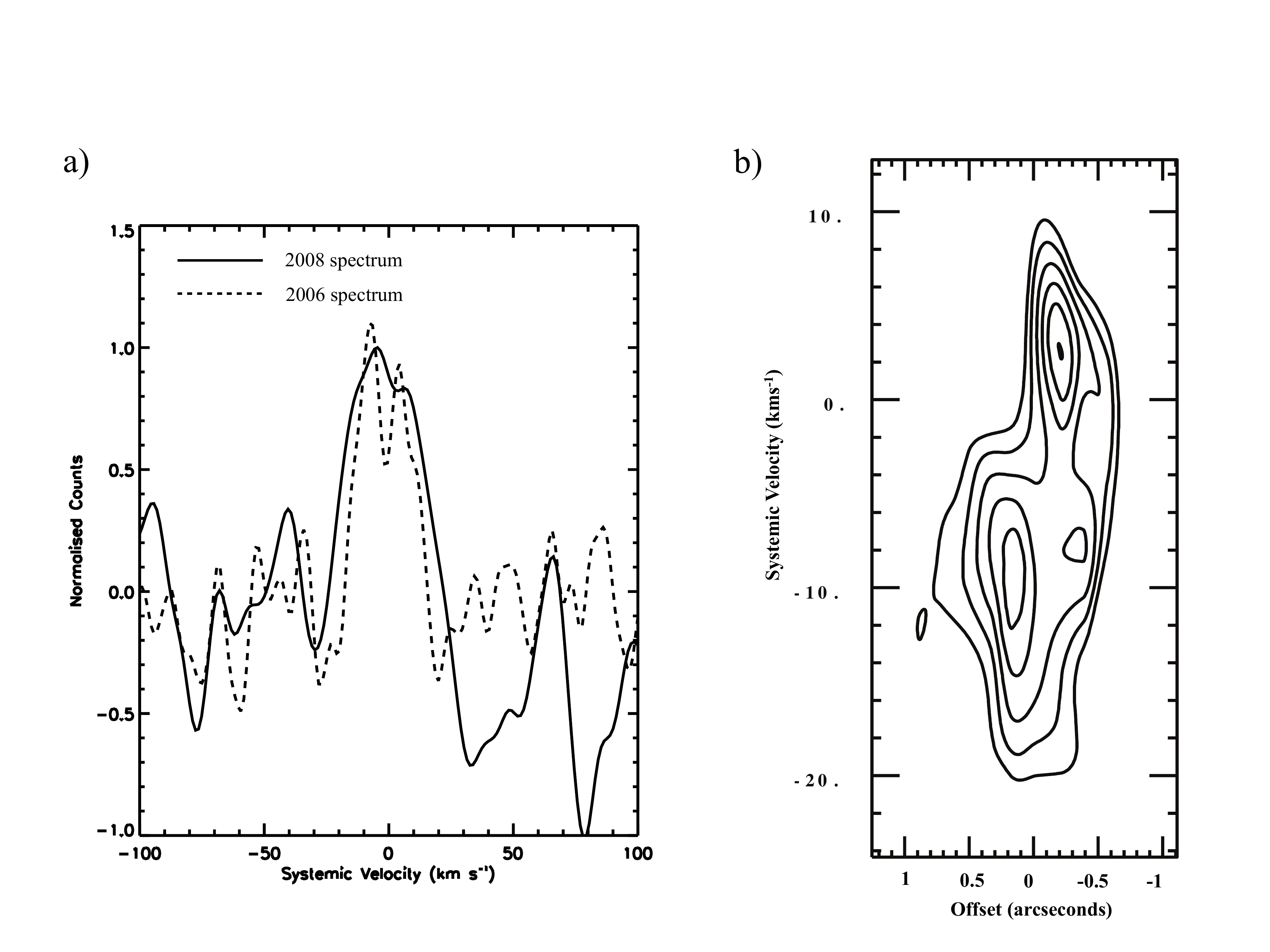}
     \caption{Left: Normalized [OI]$\lambda$6300 line profile from the 2008 (solid line) and 2006 (dashed line) spectra. The line region has been smoothed to increase the S/N, as outlined in \cite{Whelan07} and in both cases the line is double peaked. Right: Position-Velocity diagram of the [OI]$\lambda$6300 line region from the 2008 spectra. The line region is spatially resolved and clearly originates in a bipolar outflow. The contours begin at 3
times the r.m.s noise and increase in intervals of the r.m.s noise. The zero spatial offset line is measured from an accurate mapping of the continuum position using SA. The [OI]$\lambda$6300 night sky line has been subtracted from the spectrum.}
     \label{comp}
\end{figure*}

\begin{figure*}
   \includegraphics[width=18cm]{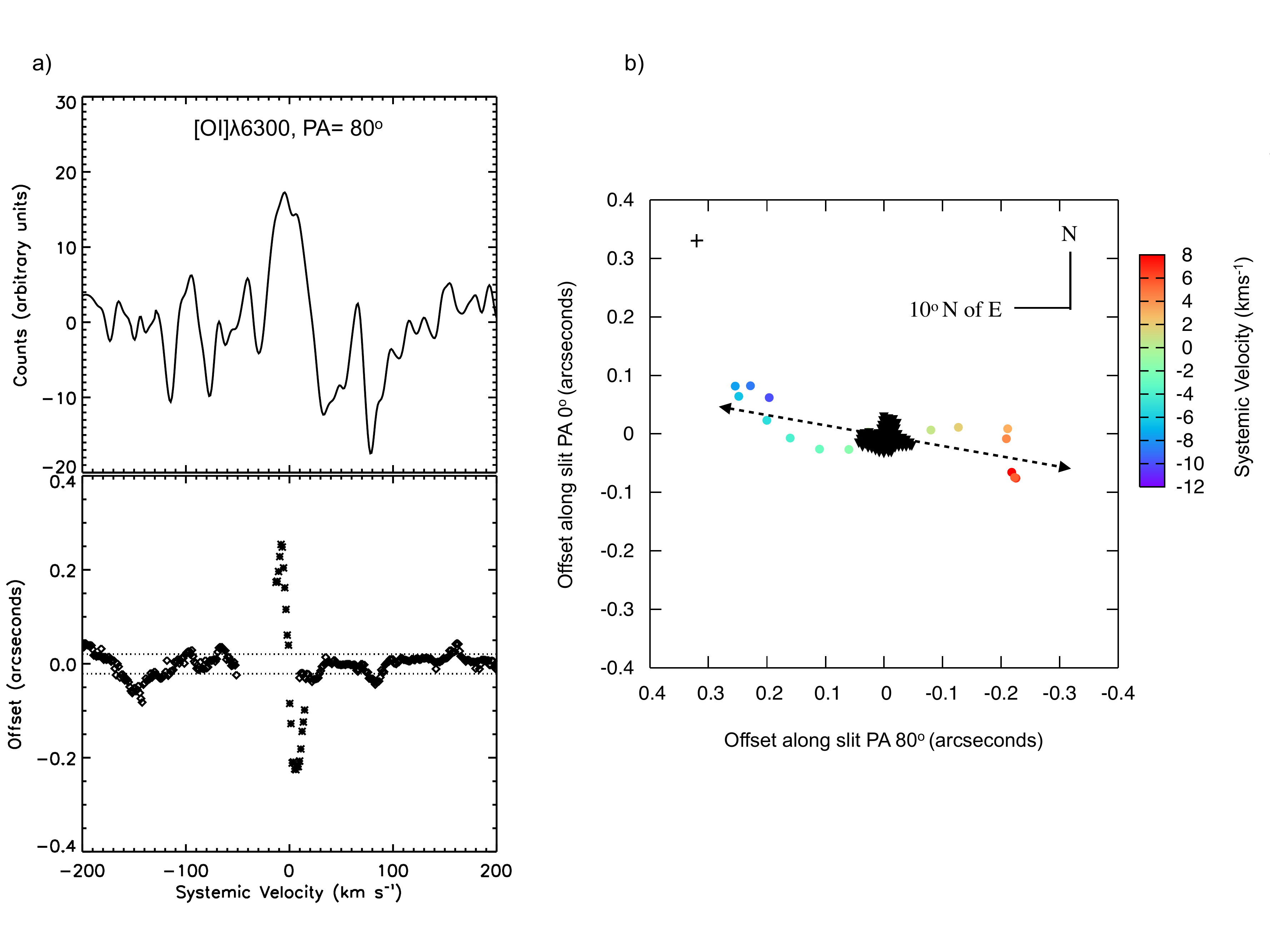}
     \caption{Left: SA of the 2008 [OI]$\lambda$6300 line. The SA clearly confirms what is revealed in the PV diagram in Figure \ref{comp}, that the blue and red peaks are offset in opposing directions. The dashed line marks the 1-$\sigma$ error in the centroid fitting. Right:  The spectro-astrometric signatures measured in the [OI]$\lambda$6300 line in 2006 and 2008 are combined to constrain the outflow PA. Note that the x and y axes do not represent orthogonal slit PAs but rather PAs of 0$^{\circ}$ and 80$^{\circ}$. }
     \label{2mass_2d}
\end{figure*}

In Figure \ref{spec1d} the 1D extracted spectrum from the UVES 2006 data is shown. This spectrum was extracted using the ESO UVES pipeline and the purpose here is to show the full extent of the UVES spectra and to mark the strongest lines. The 2006 UVES spectrum is shown as the 2008 spectrum does not include the region of the H$\beta$ line. In this study SA is applied to the [OI]$\lambda$6300, H$\alpha$ and H$\beta$ lines to search for evidence of the outflow and a comparison is made between the 2006 and 2008 spectra. 
Beginning with the [OI]$\lambda$6300 line, Figure \ref{comp} (a) compares the line profiles of the [OI]$\lambda$6300 emission in the 2006 data and in the 2008 data. In both cases the line is double peaked with the blue and red-shifted peaks at $\sim$ -8~\km\ and +4~\km, respectively. The small radial velocities of the peaks agree with the hypothesis that 2M1207A has a near-edge-on accretion disk and, therefore, an outflow close to the plane of the sky. The inclination of the 2M1207A disk has recently been estimated by \cite{Riaz12, Riaz12err} at 57$^{\circ}$ to 69$^{\circ}$. The kinematics of the outflow and spectro-astrometric results would place the inclination at the upper end of this scale. In the 2006 spectra, the emission peaks were found to be offset in opposite directions by $\sim$ 80~mas confirming the origin of the emission in an outflow. In Figure  \ref{comp} (b) the position velocity diagram of the 2008 [OI]$\lambda$6300 emission region is shown. The two peaks are clearly seen and, interestingly, they are spatially offset in opposite directions, verifying the conclusion of \cite{Whelan07} and revealing that the PA of the 2M1207A outflow lies closer to a PA of 90$^{\circ}$ than 0$^{\circ}$. This is the first observation where a BD FEL region is spatially resolved and therefore where SA is not necessary to demonstrate origin in an outflow. In Figure \ref{2mass_2d} the PA of the outflow is recovered by plotting the offsets at 0$^{\circ}$ against the offsets at 80$^{\circ}$ as a function of systemic velocity. Taking into account the fact that the two slit PAs were not perpendicular the outflow PA is constrained at $\sim$ 65$^{\circ}$ $\pm$ 10$^{\circ}$. Assuming that the outflow is perpendicular to the disk this gives a likely disk PA of $\sim$ 155$^{\circ}$. The PA of the companion is 125.6$^{\circ}$ $\pm$ 0.7$^{\circ}$ \citep{Mohanty07}. The difference of $\sim$ 30$^{\circ}$ between the disk PA (inferred from the outflow PA) and the PA of the companion could be accounted for if the companion and disk are not coplanar \citep{Riaz12, Riaz12err}. Alternatively, projection effects could explain the difference. The difference between the outflow axis projected on the sky and the PA of the planet will be 90$^{\circ}$ only at the time of the greatest elongations. Most of the rest of the time it will be near but not exactly 90$^{\circ}$ for a close to, but not exactly, edge-on orbital plane, depending on the position on the orbit.

\begin{figure*}
   \includegraphics[width=21cm]{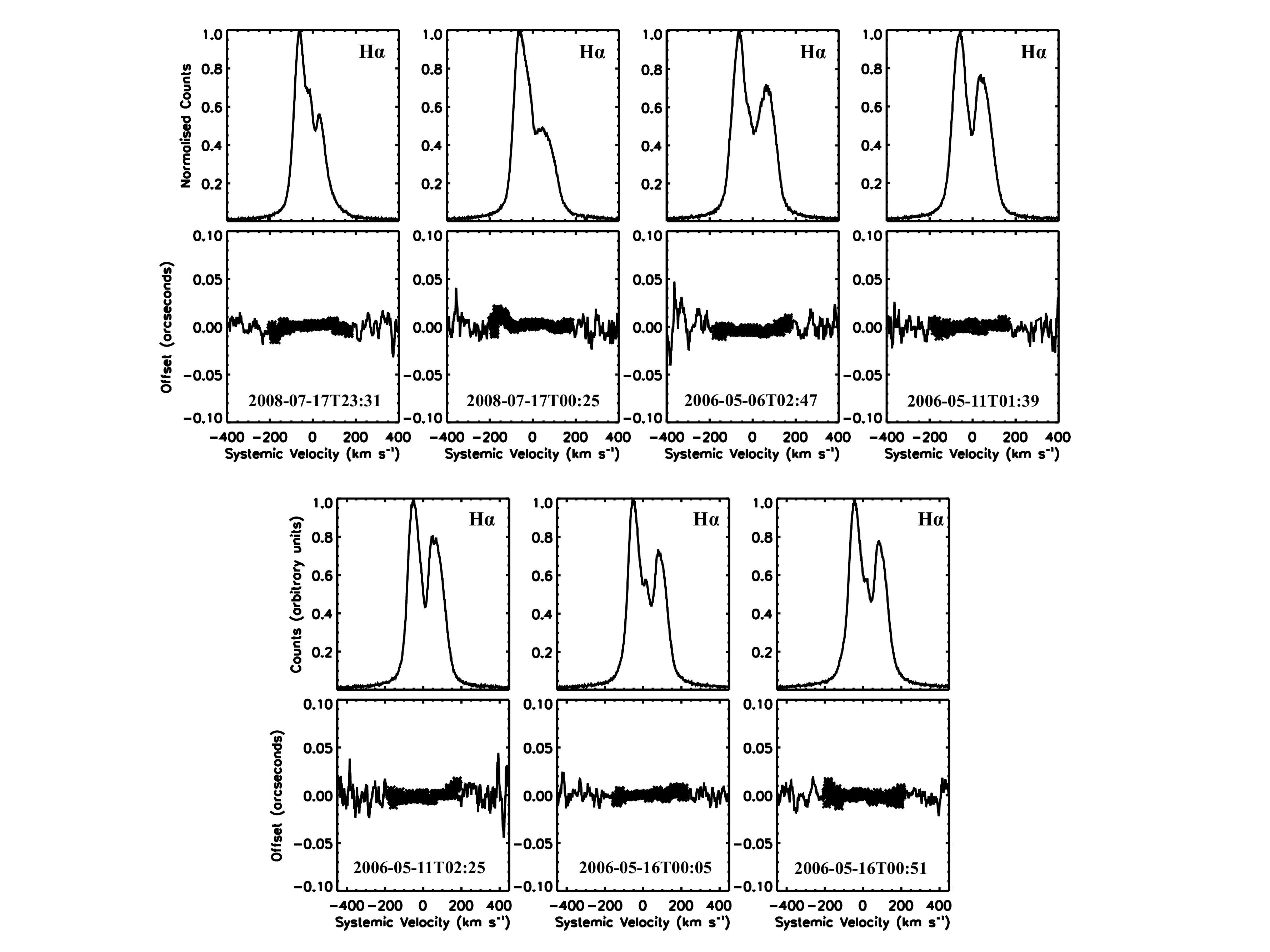}
     \caption{Spectro-astrometric analysis of the \Ha\ emission in all the UVES spectra obtained for the outflow study. The purpose here was to check if an outflow component could be detected as the \Ha\ line varied. No offset is detected in any of the spectra hence it can be concluded that the vast majority of the \Ha\ emission is likely tracing accretion in the BD and that the \Ha\ line can be used to provide a reasonable estimate of the mass accretion rate. Note that the line is highly variable as outlined by \cite{Scholz05}. The appearance of a small bump between the primary and secondary peaks was not previously detected in any variability study.}
     \label{Halpha}
\end{figure*}

\begin{figure*}
   \includegraphics[width=20cm]{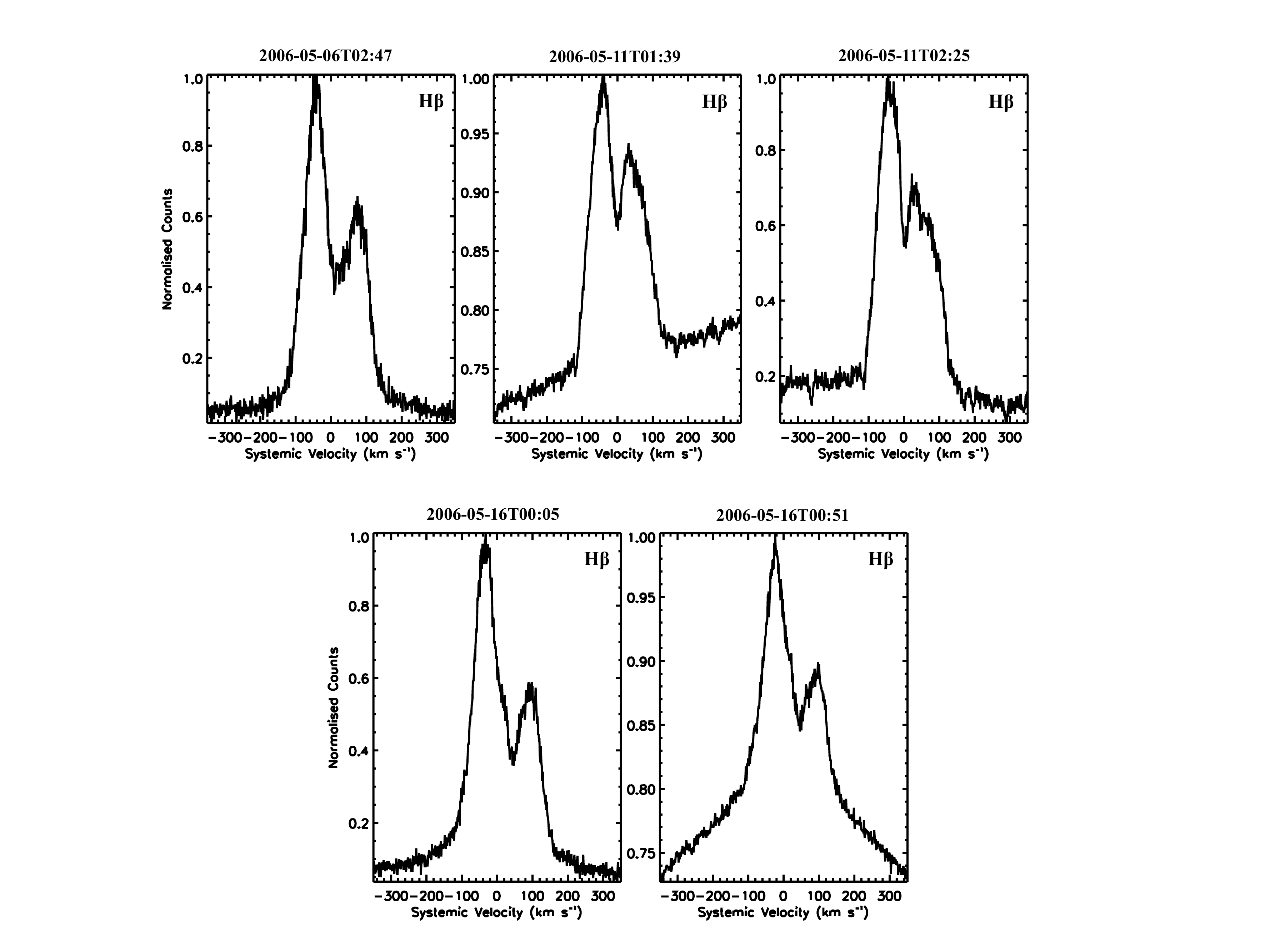}
     \caption{Comparing the H$\beta$ line profiles in the 2006 spectra. Both the line and continuum vary. No spectro-astrometric signal was detected for the H$\beta$ line.}
     \label{Hbeta}
\end{figure*}

In addition to FELs, permitted emission lines are also strong tracers of outflow activity in BDs and low mass stars \citep{Whelan09a, Whelan04}. Indeed H$\alpha$ and H$\beta$ emission is known to trace outflows and collimated jets \citep{Bacciotti2011}. Typically it is found that both outflow and accretion process contribute to the permitted line regions with the wing-emission often tracing the outflow component \citep{Podio08, Whelan04, Takami01}. SA is also used here to search for any contribution to the H$\alpha$ and H$\beta$ lines from the 2M1207A outflow. \cite{Whelan07} analyzed the 2006 H$\alpha$ emission region and no evidence of an outflow component was found. The H$\alpha$ and H$\beta$ lines of 2M1207A are, however, known to be highly variable \citep{Stelzer07, Scholz05}, and therefore one possibility that is important to consider, is that the 2006 spectra were taken during a period of high accretion activity, making it very difficult to detect any outflow component. If spectra are taken during a period of lower accretion activity the ratio between the emission from the outflow and the emission from the accretion could be more conducive to the detection of the outflow component. 
With this in mind, all the H$\alpha$ emission from all the spectra taken in 2006 and 2008 were analyzed using SA. For the analysis of the [OI]$\lambda$6300 line the spectra were median combined to increase the S/N in the [OI]$\lambda$6300 line region. However, the S/N in the region of the H$\alpha$ line is high enough to analyse the spectra separately. Five individual spectra were obtained in 2006 with a further two in 2008. Figure \ref{Halpha} compares the H$\alpha$ line profile shape and spectro-astrometric signature  in all seven spectra. While the shape changes a lot (note the appearance of a small ``bump" between the two main peaks),  no spectro-astrometric signature is detected. Hence it can be argued that the majority of the emission is coming from close to the source, presumably from accretion, and that the H$\alpha$ should provide a good estimate of the mass accretion rate of 2M1207A.  The small ``bump" is interesting as it was not previously detected in the variability studies of \cite{Stelzer07} or \cite{Scholz05}. Similarly we compare the line profile shape of the H$\beta$ and spectra-astrometric signature of the H$\beta$ line. Again no spectro-astrometric signal is detected and only the line profiles are presented in Figure \ref{Hbeta}.

\subsection{[SII] Imaging: Elongated PSF}
\label{fors1}
\begin{figure*}
   \includegraphics[width=19cm]{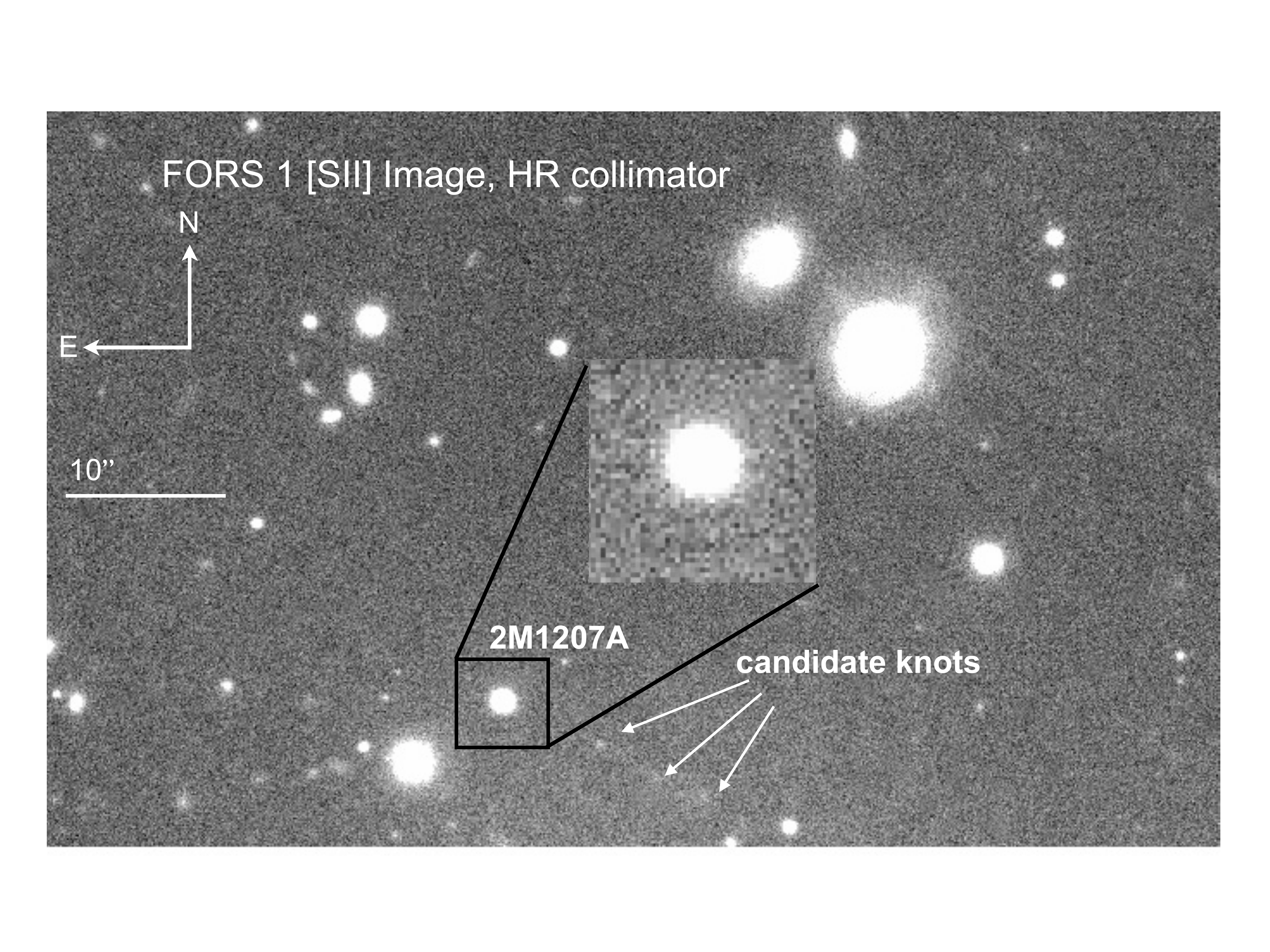}
     \caption{[SII] image in the region of 2M1207. A zoom on 2M1207 (the A and B components to the system are not resolved) is shown in an inset. Note the slight elongation in the PSF along a similar PA to the outflow PA derived from the spectro-astrometric analysis. The PA of 2M1207B with respect to 2M1207A is 125$^{\circ}$. The position of the possible outflow features are marked. As described in the text two images with exposure times of 1800s were obtained. The image with the better seeing is shown here.}
     \label{imageSII}
\end{figure*}

The FORS-1 [SII] image with an inset of a zoom on 2M1207A is shown in Figure \ref{imageSII}. Firstly we chose to look for evidence of emission from the outflow close to the BD by analyzing the image with the best seeing ($\sim$ 0\farcs6). Interestingly, the PSF appears to be extended along the outflow PA suggested by the spectro-astrometric results. In Figure \ref{image_comp} the PSF of 2M1207A in the [SII] image is compared with the PSF of two nearby stars in [SII] and with 2M1207A in R. Each PSF was fitted with an ellipse and the results compared. The 2M1207A PSF is more elliptical than the others and is elongated along a PA given by the elliptical fit at 62$^{\circ}$ $\pm$ 6$^{\circ}$. This is consistent with the estimate of 65$^{\circ}$ $\pm$ 10$^{\circ}$ for the outflow PA. This agreement, along with the fact that the extension is only seen in the [SII] PSF of 2M1207A, is strong evidence against an artifact being the cause of the extension. Hence, we argue that we are in fact detecting the 2M1207A outflow in the [SII] image. However, note that no [SII] emission is detected at the source position in the UVES spectra (see Figure 2). Why [SII] outflow emission should be detected in the FORS1 image and not in the UVES spectra requires further investigation. Using 1d flux calibrated spectra from the UVES pipeline, the flux of the 2M1207A [OI]$\lambda$6300 emission line is estimated at 1 $\times$ 10$^{-16}$ erg/cm$^{2}$/s and the continuum flux included in the range of the [SII] narrowband filter at 37 $\times$ 10$^{-16}$ erg/cm$^{2}$/s (0.6 $\times$ 10$^{-16}$ erg/cm$^{2}$/s/\AA). Using UVES spectra of other BD/VLMS where both [OI] and [SII] emission is detected we measured the average flux ratio of the [SII]$\lambda$6716 / [OI]$\lambda$6300 and [SII]$\lambda$6731 / [OI]$\lambda$6300 lines and use these values to make a realistic estimate of the ratio between the [SII] and [OI] emission in 2M1207A. If this same ratio holds for 2M1207A, then the [SII] lines would have strengths of 0.18 and 0.3 $\times$ 10$^{-16}$ erg/cm$^{2}$/s, respectively. This puts the [SII] line at around the 1-$\sigma$ to 2-$\sigma$ detection level, explaining its non-detection above the continuum. 

Due to the difference in critical density between the [OI]$\lambda$6300 and [SII]$\lambda\lambda$6716, 6731 lines, the [OI] emission is stronger than the [SII] emission at the source position and the [SII] emission is normally more extended. The critical densities of [OI]$\lambda$6300 and [SII]$\lambda$6731 lines, at T $\sim$ 8,000~K, can be taken as 2 $\times$ 10$^{6}$~cm$^{-3}$ and 1.3 $\times$ 10$^{4}$~cm$^{-3}$ respectively \citep{Hartigan1995}. Thus, knots further out in a jet are brighter in [SII] than [OI] and this is illustrated, for example, for the Par-Lup3-4 outflow in \cite{Bacciotti2011}. To check for faint extended [SII] emission we summed the spectra over $\pm$ 1\arcsec\ and the results are plotted in Figure 9. Indeed, once the spectra are summed in the spatial direction, both [SII] lines are detected at a velocity comparable to the [OI]$\lambda$6300 velocity. The spectra have been smoothed in the same way as the [OI]$\lambda$6300 spectrum and also shown is the Li I photospheric line which is used to measure the systemic velocity. Spectra from 2006 and 2008 are compared and  in both cases the [SII]$\lambda$6716 line is brighter than the [SII]$\lambda$6731 line indicating a low value of n$_{e}$ of  $\sim$ 1000~cm$^{-3}$ \citep{Osterbrock88}. The fact that some extended [SII] emission is detected in the UVES spectra supports our detection of the outflow in the image.




\subsection{[SII] Imaging: Extended Emission Features}

To the south-west of 2M1207A, a series of faint knot-like features which culminate in a bow-shock shaped object at a distance of $\sim$ 16~\arcsec\ are detected (see Figures 7 and 10). These will now be referred to as features {\it A} to {\it D}, with {\it D} being the feature furthest from the BD. These sources seen in [SII], R and I  are in the direction of the red-shifted outflow and look very much ``to the eye" like they could be outflow knots.  They are well fitted with a PA of 245$^{\circ}$ as expected for the red-shifted flow. However, it is possible that {\it A - D} could be faint stars or galaxies and therefore further analysis is needed to confirm that they are HH objects forming part of the 2M1207A outflow. Firstly a [SII]$-$I-band subtraction was carried out  to isolate any outflow emission. In Figure 10 the FORS1 [SII], R and I-band images and a SUSI2 I-band image downloaded from the ESO archive are compared. The features are clearly detected in all bands. The R-BESS filter covers the range 5820-7320~\AA\ and the I-BESS filter the range 6990-8370~\AA. Hence, both filters include outflow tracers, especially the R-BESS which covers strong tracers like [OI]$\lambda$6300, \Ha, [NII]$\lambda$6583 and [SII]$\lambda$6731. Outflow tracers such as [FeII]$\lambda$7155 and [CaII]$\lambda$7291 are found in the range of the I-band. However, as these lines are fainter than the lines in the range of the R-band and sometimes not present, the outflow emission in the I-band is generally significantly less than, for example, that of the [SII] narrowband filter \citep{Mundt83, B81}. Therefore, I-band broadband images are routinely used to remove continuum emission from a narrowband image and thus to enhance any outflow emission. 
In Figure \ref{smoothed} a Gaussian smoothed FORS1 [SII]$-$I image of the region where it was possible to remove the I-band fringing is shown.
The emission regions {\it A-D} are detected and the bow-shaped nature of feature {\it D} is obvious. The red line marks the position of outflow PA which well fits knot {\it D} in particular. The emission regions {\it B} and {\it C} are just above the noise in the [SII]$-$I image while {\it A} is the brightest.  
 
To further eliminate the possibility that {\it A-D} are stars / galaxies a [SII]$-$I versus R$-$I color-color analysis was carried out and is presented in Figure \ref{color}. 
A HH object is expected to display a blue broadband R-I color due to 
the contribution of the Halpha and [SII] 6716, 6731 lines to the R band, 
and an even bluer [SII]-I color caused by the larger fraction of the 
[SII] filter passband in which [SII] emission is noticeable, in contrast 
with the redder R-I and [SII]-I colors of continuum-dominated sources 
like stars and normal galaxies.
Therefore this analysis should separate any HH objects from stars and galaxies.
Firstly synthetic colors for stars with T$_{eff}$ in the range 20000 ~K to 1500~K are plotted (blue squares). These were calculated for the wavelength ranges of the FORS1 filters using a simple blackbody model. The hottest stars lie close to the origin while the coolest BDs lie in the tail at the end of the sequence (upper right). The direction of the reddening is also plotted. In this figure the colors for both the TW Hya and Cha I star forming regions (based on the FOV of our images) are also plotted. The idea is to compare the colors of the sources in our images of the 2M1207 region not only to the synthetic colors but also to another star forming region and any known HH objects within that region. Cha I was chosen as suitable images were available in the ESO archive which included the protostar ESO-H$\alpha$ 574 and its outflow  \citep{Bacciotti2011}. The black triangles represent the sources in the FOV of the TW Hya images and the black asterisks the sources in Cha I. For both star forming regions the sources follow the same sequence as the synthetic colors with the hottest objects clustering around the origin and the coolest in the tail.

Looking first at the TW Hya data, the purple triangles are galaxies that could be identified in the field and the green triangles represent features {\it A} and {\it D}. {\it B} and {\it C} were omitted as they were too faint for their colors to be measured. While {\it A} lies amongst the stars and galaxies ruling out the possibility that it is a HH object, {\it D} is offset from the stellar sequence and is bluer than the stars and galaxies plotted. Thus considering the bluer nature of {\it D}, that it lies along the outflow PA and that it is bow-shaped with the bow pointing away from the BD, we conclude that it is a HH object in the 2M1207A outflow. This conclusion is further strengthened by comparison with the knots in the ESO-H$\alpha$ 574 outflow. ESO-H$\alpha$ 574 powers a well-developed bipolar jet first observed by \cite{Comeron06}. \cite{Comeron06} identified for knots in the blue-shifted lobe of the ESO-H$\alpha$ 574 flow which they also labelled A-D and Knot E in the red-shifted flow. Here we label them HA574-A to HA574-D, to distinguish them from the features in the 2M1207A flow. In the images HA574-E was the only knot for which the colors could be measured as the others were not well enough separated and too faint in the I band. Its color is represented in Figure 12 by the purple asterisk. HA574-E is also bluer and well separated from the stellar sequence as expected. The fact that it is bluer than {\it D} can be explained by the relatively weak [SII]$\lambda\lambda$6716, 6731 emission from {\it D} compared to HA574-E. Flux calibrated spectra of the ESO-H$\alpha$ 574 outflow which included  HA574-A,  HA574-B and  HA574-E were also available (J. Alcala, priv. comm, also see Bacciotti et al. 2011). These spectra were used to estimate the colors for these three knots (purple squares). The color for HA574-E lies close to that estimated from the images while there is a spread in the colors of HA574-A and HA574-B, again consistent with differences in shock conditions. 

\section{Discussion and Conclusions}

Of the small number of BD outflows studied to date, 2M1207A is perhaps the most interesting due to its low mass, age of some 10~Myr and binarity.  With the new results presented here we add considerably to our 2007 paper reporting the discovery of this outflow. The 2008 spectra confirm the original results, resolve for the first time for a BD the optical outflow, and allow an estimate of the outflow PA to be made. Furthermore, the outflow is detected in the [SII] image in the form of an elongation in the PSF and at least one HH object (feature {\it D}). Evidence in support of the extended features {\it B-D} being part of the 2M1207A outflow include: the fact that they lie along the estimated outflow PA; that the terminal feature {\it D}, is bow-shock shaped with the apex pointing away from the source and that {\it D} is separated in Figure \ref{color} from the stars and galaxies. While the color-color analysis shown in Figure 12 demonstrates that at least {\it D} is a probable HH object, it shows that {\it A} is a star or galaxy.  New higher S/N images would greatly add to current understanding of this outflow, as firstly one would be able to study the features with a much improved S/N and secondly a proper motion study, which would confirm the origin of {\it B, C}, could be conducted. Using the upper estimate of the disk inclination angle  given by \cite{Riaz12err, Riaz12} (69$^{\circ}$) and combining it with the measured radial velocities, we estimate the tangential velocity of the 2M1207A outflow at $\sim$ 20~kms$^{-1}$. While at the lower end of the range, this is in line with previous estimates of tangential velocities for a sample of BD outflows given by \cite{Whelan09b}. 
Also the non-detection of a
spectroastrometric signal for H$\alpha$ is consistent with a low 
tangential velocity as with such 
low velocity one would expect only weak shocks and hardly any ionization 
of H in the jet and therefore no H$\alpha$. A velocity of $\sim$ 20 kms$^{-1}$ would give feature {\it D} an age of some 173 years. Taking that any new images of 2M1207A would be observed in January  2013 at the earliest, this would mean a $\sim$ 57 month time difference between these and the 2008 images. Thus a shift of at least $\sim$ 0\farcs5 in features {\it B-D} would be expected, if they are indeed part of the outflow. Such proper motions should be easily measured. The SUSI2 images shown in Figure \ref{extended} were obtained in January 2006, meaning that a shift of $\sim$ 0\farcs2, could reasonably be expected in the knots in the I band. Attempts to measure such a proper motion between these two sets of I-band images were inconclusive. A high quality spectrum along the PA of the outflow to search for shock related emission lines would also be of great benefit. New imaging and spectroscopic observations would also enable a search for emission features along the blue-shifted outflow to be conducted. Finally we note that while it is normally the case that the blue-shifted lobe is preferentially detected, there are cases where the red-shifted flow is more prominent \citep{Whelan09b, Woitas02}. 

Results presented here also demonstrate that the 2M1207A outflow is well-collimated. This confirms that BD outflows can be collimated like low mass protostellar jets, further adding to the properties they share \citep{Whelan09b}. As BD optical outflows had not been directly spatially resolved previous to this work, the extent of their collimation was uncertain.  However, the detection of a molecular outflow from ISO-Oph~102 by \cite{Phan2008} did point to some degree of focussing, as it is a collimated jet which is postulated to be driving the molecular outflows from low mass stars \citep{Bach96}. Measurements of jet width as a function of distance along the jet for classical T Tauri stars (CTTS) have shown that collimation occurs on scales of a few tens of AU from the star \citep{Ray2007} which is consistent with magneto-hydrodynamical models \citep{Dougados09}. As well as being collimated, jets from low mass stars are also often episodic in that they are made up of a series of distinct knots assumed to be associated with separate accretion events. When discussing episodic outflows the striking H$_{2}$ flow known as HH~212, driven by a Class 0 young low mass star, immediately comes to mind \citep{Zinnecker98}. The HH~212 flow is very well defined with matching pairs of bow-shock shaped knots on either side of the driving source. By studying the individual knots in protostellar jets it is possible to get a picture of the mass loss and accretion history of the source. Our study strongly indicates that accretion and outflow activity proceeds in this extremely low mass object in an analogous way to low mass protostars, and by studying this flow we will be able to investigate the history of the 2M1207 system.

As the current sample of known BD outflows is small the present aim of our study of BD outflow activity is to greatly increase the sample. The approach of using SA to identify BD outflows, while effective, is slow as high quality, high S/N spectra are required and thus, observations with an 8~m class telescope or better are generally necessary. In addition, if one wants to thoroughly investigate the properties of the outflow, including the morphology, kinematics and mass loss rate, spectra along the outflow PA are needed. Hence, to identify the outflow sources and constrain the outflow PA, two spectra at orthogonal PAs should be observed and then a follow-up spectrum must be taken along the estimated PA.  We consider that the imaging observations of the 2M1207A outflow presented here, offer an underutilized  method way of identify BD outflows. A [SII] imaging study to look for extensions in the PSFs of BD candidates would be a fast way of identifying outflows as more than one source could be observed at the same time and we would expect to reach a better S/N in any [SII] outflow emission with an imager than a spectrograph for comparable exposure times. A serious limitation to imaging studies of jets from low mass protostars is the strength of the continuum emission which is many times stronger than the jet emission. For this reason the jet is often not seen in the image until several arc-seconds from the source. As the continuum emission in the vicinity of the [SII]$\lambda\lambda$6716, 6731 lines in BDs and VLMSs is fainter than in CTTSs, this enhances the possibility of detecting any outflows in an imaging study.

Furthermore, a significant number of the BDs / VLMSs  shown spectroscopically to have outflows (4 out of 6) also have near-edge-on disks  \citep{Bacciotti2011, Whelan09b, Whelan09a} . The effect of this is to obscure the emission from the BD or star and further enhance the contrast between the jet and continuum emission in favor of the jet emission. For example, the 100~M$_{JUP}$ mass object Par-Lup3-4 is known to be under-luminous and an examination of archived FORS1 images and UVES spectra allows the ratio of the [SII] jet emission to continuum emission in the range of the [SII]+62 filter to be estimated at 50$\%$. Thus, the Par-Lup3-4 jet is easily detected in the FORS image \citep{Fernandez05}. We are confident of repeating the imaging observations of 2M1207A discussed here for other BDs and especially those with edge-on disks. Such observations would of course immediately give the jet PA allowing for a more detailed follow-up study of accretion and outflow activity from high quality spectra.


\bibliographystyle{aa}

\bibliography{references}

\begin{thebibliography}{41}
\expandafter\ifx\csname natexlab\endcsname\relax\def\natexlab#1{#1}\fi

\bibitem[{{Appenzeller} {et~al.}(2000){Appenzeller}, {Bender}, {B{\"o}hnhardt},
  {Cristiani}, {Dietrich}, {Fricke}, {F{\"u}rtig}, {G{\"a}ssler}, {Gilmozzi},
  {H{\"a}fner}, {Harke}, {Heidt}, {Hess}, {Hopp}, {Hummel}, {J{\"a}ger},
  {J{\"u}rgens}, {Kudritzki}, {K{\"u}mmel}, {Mantel}, {Mehlert}, {Meisl},
  {Moellenhoff}, {Muschielok}, {Nicklas}, {Renzini}, {Rosati}, {Rupprecht},
  {Saglia}, {Seifert}, {Seitz}, {Spyromilio}, {Stahl}, {Szeifert}, \&
  {Tarantik}}]{App00}
{Appenzeller}, I., {Bender}, R., {B{\"o}hnhardt}, H., {et~al.} 2000, in From
  Extrasolar Planets to Cosmology: The VLT Opening Symposium, ed. {J.~Bergeron
  \& A.~Renzini}, 3--540

\bibitem[{{Bacciotti} {et~al.}(2011){Bacciotti}, {Whelan}, {Alcal{\'a}},
  {Nisini}, {Podio}, {Randich}, {Stelzer}, \& {Cupani}}]{Bacciotti2011}
{Bacciotti}, F., {Whelan}, E.~T., {Alcal{\'a}}, J.~M., {et~al.} 2011, \apjl,
  737, L26

\bibitem[{{Bachiller}(1996)}]{Bach96}
{Bachiller}, R. 1996, \araa, 34, 111

\bibitem[{{Barrado Y Navascu{\'e}s}(2006)}]{Barrodo06}
{Barrado Y Navascu{\'e}s}, D. 2006, \aap, 459, 511

\bibitem[{{Brugel} {et~al.}(1981){Brugel}, {Boehm}, \& {Mannery}}]{B81}
{Brugel}, E.~W., {Boehm}, K.~H., \& {Mannery}, E. 1981, \apjs, 47, 117

\bibitem[{{Chauvin} {et~al.}(2004){Chauvin}, {Lagrange}, {Dumas}, {Zuckerman},
  {Mouillet}, {Song}, {Beuzit}, \& {Lowrance}}]{Chauvin04}
{Chauvin}, G., {Lagrange}, A.-M., {Dumas}, C., {et~al.} 2004, \aap, 425, L29

\bibitem[{{Comer{\'o}n} \& {Fern{\'a}ndez}(2011)}]{Comeron11}
{Comer{\'o}n}, F. \& {Fern{\'a}ndez}, M. 2011, \aap, 528, A99

\bibitem[{{Comer{\'o}n} {et~al.}(2003){Comer{\'o}n}, {Fern{\'a}ndez},
  {Baraffe}, {Neuh{\"a}user}, \& {Kaas}}]{Comeron03}
{Comer{\'o}n}, F., {Fern{\'a}ndez}, M., {Baraffe}, I., {Neuh{\"a}user}, R., \&
  {Kaas}, A.~A. 2003, \aap, 406, 1001

\bibitem[{{Comer{\'o}n} \& {Reipurth}(2006)}]{Comeron06}
{Comer{\'o}n}, F. \& {Reipurth}, B. 2006, \aap, 458, L21

\bibitem[{{Dekker} {et~al.}(2000){Dekker}, {D'Odorico}, {Kaufer}, {Delabre}, \&
  {Kotzlowski}}]{Dekker00}
{Dekker}, H., {D'Odorico}, S., {Kaufer}, A., {Delabre}, B., \& {Kotzlowski}, H.
  2000, in Society of Photo-Optical Instrumentation Engineers (SPIE) Conference
  Series, Vol. 4008, Society of Photo-Optical Instrumentation Engineers (SPIE)
  Conference Series, ed. {M.~Iye \& A.~F.~Moorwood}, 534--545

\bibitem[{{Dougados}(2009)}]{Dougados09}
{Dougados}, C. 2009, in SF2A-2009: Proceedings of the Annual meeting of the
  French Society of Astronomy and Astrophysics, ed. M.~{Heydari-Malayeri},
  C.~{Reyl'E}, \& R.~{Samadi}, 3

\bibitem[{{Fern{\'a}ndez} \& {Comer{\'o}n}(2001)}]{Fernandez01}
{Fern{\'a}ndez}, M. \& {Comer{\'o}n}, F. 2001, \aap, 380, 264

\bibitem[{{Fern{\'a}ndez} \& {Comer{\'o}n}(2005)}]{Fernandez05}
{Fern{\'a}ndez}, M. \& {Comer{\'o}n}, F. 2005, \aap, 440, 1119

\bibitem[{{Ferreira} {et~al.}(2006){Ferreira}, {Dougados}, \&
  {Cabrit}}]{Ferreira2006}
{Ferreira}, J., {Dougados}, C., \& {Cabrit}, S. 2006, \aap, 453, 785

\bibitem[{{Hartigan} {et~al.}(1995){Hartigan}, {Edwards}, \&
  {Ghandour}}]{Hartigan1995}
{Hartigan}, P., {Edwards}, S., \& {Ghandour}, L. 1995, \apj, 452, 736

\bibitem[{{Machida} {et~al.}(2009){Machida}, {Inutsuka}, \&
  {Matsumoto}}]{Machida09}
{Machida}, M.~N., {Inutsuka}, S.-i., \& {Matsumoto}, T. 2009, \apjl, 699, L157

\bibitem[{{Mohanty} {et~al.}(2003){Mohanty}, {Jayawardhana}, \& {Barrado y
  Navascu{\'e}s}}]{Mohanty03}
{Mohanty}, S., {Jayawardhana}, R., \& {Barrado y Navascu{\'e}s}, D. 2003,
  \apjl, 593, L109

\bibitem[{{Mohanty} {et~al.}(2005){Mohanty}, {Jayawardhana}, \&
  {Basri}}]{Mohanty05}
{Mohanty}, S., {Jayawardhana}, R., \& {Basri}, G. 2005, \apj, 626, 498

\bibitem[{{Mohanty} {et~al.}(2007){Mohanty}, {Jayawardhana}, {Hu{\'e}lamo}, \&
  {Mamajek}}]{Mohanty07}
{Mohanty}, S., {Jayawardhana}, R., {Hu{\'e}lamo}, N., \& {Mamajek}, E. 2007,
  \apj, 657, 1064

\bibitem[{{Mundt} {et~al.}(1983){Mundt}, {Stocke}, \& {Stockman}}]{Mundt83}
{Mundt}, R., {Stocke}, J., \& {Stockman}, H.~S. 1983, \apjl, 265, L71

\bibitem[{{Osterbrock}(1988)}]{Osterbrock88}
{Osterbrock}, D.~E. 1988, \pasp, 100, 412

\bibitem[{{Phan-Bao} {et~al.}(2008){Phan-Bao}, {Riaz}, {Lee}, {Tang}, {Ho},
  {Mart{\'{\i}}n}, {Lim}, {Ohashi}, \& {Shang}}]{Phan2008}
{Phan-Bao}, N., {Riaz}, B., {Lee}, C.-F., {et~al.} 2008, \apjl, 689, L141

\bibitem[{{Podio} {et~al.}(2008){Podio}, {Garcia}, {Bacciotti}, {Antoniucci},
  {Nisini}, {Dougados}, \& {Takami}}]{Podio08}
{Podio}, L., {Garcia}, P.~J.~V., {Bacciotti}, F., {et~al.} 2008, \aap, 480, 421

\bibitem[{{Ray} {et~al.}(2007){Ray}, {Dougados}, {Bacciotti}, {Eisl{\"o}ffel},
  \& {Chrysostomou}}]{Ray2007}
{Ray}, T., {Dougados}, C., {Bacciotti}, F., {Eisl{\"o}ffel}, J., \&
  {Chrysostomou}, A. 2007, Protostars and Planets V, 231

\bibitem[{{Riaz} \& {Gizis}(2008)}]{Riaz08}
{Riaz}, B. \& {Gizis}, J.~E. 2008, \apj, 681, 1584

\bibitem[{{Riaz} {et~al.}(2012{\natexlab{a}}){Riaz}, {Lodato}, {Stamatellos},
  \& {Gizis}}]{Riaz12err}
{Riaz}, B., {Lodato}, G., {Stamatellos}, D., \& {Gizis}, J.~E.
  2012{\natexlab{a}}, \mnras, 424, L74

\bibitem[{{Riaz} {et~al.}(2012{\natexlab{b}}){Riaz}, {Lodato}, {Stamatellos},
  \& {Gizis}}]{Riaz12}
{Riaz}, B., {Lodato}, G., {Stamatellos}, D., \& {Gizis}, J.~E.
  2012{\natexlab{b}}, \mnras, 422, L6

\bibitem[{{Scholz} {et~al.}(2005){Scholz}, {Jayawardhana}, \&
  {Brandeker}}]{Scholz05}
{Scholz}, A., {Jayawardhana}, R., \& {Brandeker}, A. 2005, \apjl, 629, L41

\bibitem[{{Stelzer} {et~al.}(2007){Stelzer}, {Scholz}, \&
  {Jayawardhana}}]{Stelzer07}
{Stelzer}, B., {Scholz}, A., \& {Jayawardhana}, R. 2007, \apj, 671, 842

\bibitem[{{Szeifert} {et~al.}(1998){Szeifert}, {Appenzeller}, {Fuertig},
  {Seifert}, {Stahl}, {Boehnhardt}, {Gaebler}, {Haefner}, {Hess}, {Mantel},
  {Meisl}, {Muschielok}, {Tarantik}, {Harke}, {Juergens}, {Nicklas}, \&
  {Rupprecht}}]{Szei98}
{Szeifert}, T., {Appenzeller}, I., {Fuertig}, W., {et~al.} 1998, in Society of
  Photo-Optical Instrumentation Engineers (SPIE) Conference Series, Vol. 3355,
  Society of Photo-Optical Instrumentation Engineers (SPIE) Conference Series,
  ed. {S.~D'Odorico}, 20--27

\bibitem[{{Takami} {et~al.}(2001){Takami}, {Bailey}, {Gledhill},
  {Chrysostomou}, \& {Hough}}]{Takami01}
{Takami}, M., {Bailey}, J., {Gledhill}, T.~M., {Chrysostomou}, A., \& {Hough},
  J.~H. 2001, \mnras, 323, 177

\bibitem[{{Wang} \& {Henning}(2006)}]{Wang06}
{Wang}, H. \& {Henning}, T. 2006, \apj, 643, 985

\bibitem[{{Whelan} \& {Garcia}(2008)}]{Whelan08}
{Whelan}, E. \& {Garcia}, P. 2008, in Lecture Notes in Physics, Berlin Springer
  Verlag, Vol. 742, Jets from Young Stars II, ed. {F.~Bacciotti, L.~Testi, \&
  E.~Whelan}, 123

\bibitem[{{Whelan} {et~al.}(2010){Whelan}, {Dougados}, {Perrin}, {Bonnefoy},
  {Bains}, {Redman}, {Ray}, {Bouy}, {Benisty}, {Bouvier}, {Chauvin}, {Garcia},
  {Grankvin}, \& {Malbet}}]{Whelan10}
{Whelan}, E.~T., {Dougados}, C., {Perrin}, M.~D., {et~al.} 2010, \apjl, 720,
  L119

\bibitem[{{Whelan} {et~al.}(2009{\natexlab{a}}){Whelan}, {Ray}, \&
  {Bacciotti}}]{Whelan09a}
{Whelan}, E.~T., {Ray}, T.~P., \& {Bacciotti}, F. 2009{\natexlab{a}}, \apjl,
  691, L106

\bibitem[{{Whelan} {et~al.}(2005){Whelan}, {Ray}, {Bacciotti}, {Natta},
  {Testi}, \& {Randich}}]{Whelan05}
{Whelan}, E.~T., {Ray}, T.~P., {Bacciotti}, F., {et~al.} 2005, \nat, 435, 652

\bibitem[{{Whelan} {et~al.}(2004){Whelan}, {Ray}, \& {Davis}}]{Whelan04}
{Whelan}, E.~T., {Ray}, T.~P., \& {Davis}, C.~J. 2004, \aap, 417, 247

\bibitem[{{Whelan} {et~al.}(2009{\natexlab{b}}){Whelan}, {Ray}, {Podio},
  {Bacciotti}, \& {Randich}}]{Whelan09b}
{Whelan}, E.~T., {Ray}, T.~P., {Podio}, L., {Bacciotti}, F., \& {Randich}, S.
  2009{\natexlab{b}}, \apj, 706, 1054

\bibitem[{{Whelan} {et~al.}(2007){Whelan}, {Ray}, {Randich}, {Bacciotti},
  {Jayawardhana}, {Testi}, {Natta}, \& {Mohanty}}]{Whelan07}
{Whelan}, E.~T., {Ray}, T.~P., {Randich}, S., {et~al.} 2007, \apjl, 659, L45

\bibitem[{{Woitas} {et~al.}(2002){Woitas}, {Ray}, {Bacciotti}, {Davis}, \&
  {Eisl{\"o}ffel}}]{Woitas02}
{Woitas}, J., {Ray}, T.~P., {Bacciotti}, F., {Davis}, C.~J., \&
  {Eisl{\"o}ffel}, J. 2002, \apj, 580, 336

\bibitem[{{Zinnecker} {et~al.}(1998){Zinnecker}, {McCaughrean}, \&
  {Rayner}}]{Zinnecker98}
{Zinnecker}, H., {McCaughrean}, M.~J., \& {Rayner}, J.~T. 1998, \nat, 394, 862

\end{thebibliography}

\acknowledgements{E.T. Whelan is supported by an IRCSET-Marie Curie International Mobility Fellowship in Science, Engineering and Technology within the 7th European Community Framework Programme; T.P. Ray acknowledges support from Science Foundation Ireland under the Research Frontiers Program grant 11/RFP/AST/3331; P.J. Kavanagh is funded through the BMBF/DLR grant 50-0R-1009. }

\begin{figure*}
   \includegraphics[width=18cm]{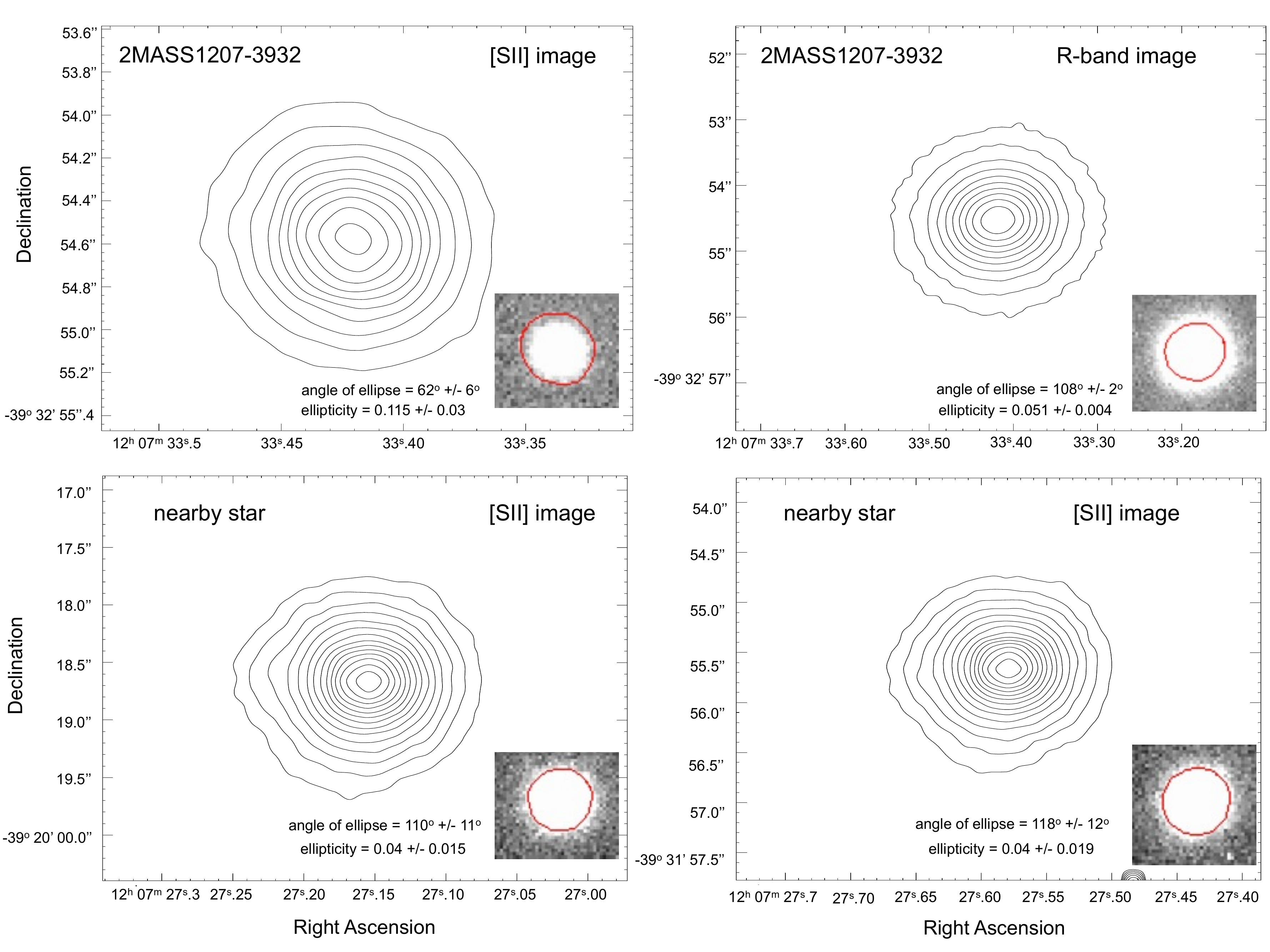}
     \caption{Comparison of the PSF of 2M1207A in [SII] with the PSFs of 2M1207A in R and the PSFs of two nearby stars in [SII]. The purpose here is to demonstrate that the 2M1207A PSF in [SII] is elongated along the outflow PA with respect to the other PSFs. An elliptical fit to the PSFs is shown in the inset. The ellipticity of the fits shows that the 2M1207 [SII] PSF is indeed more elliptical and has a PA estimated from the fit at 62$^{\circ}$ $\pm$ 6$^{\circ}$. This PA is compatible with the outflow PA as estimated using SA. We use this comparison as strong evidence that we are detecting the 2M1207A outflow in the [SII] image. }
     \label{image_comp}
\end{figure*}

\begin{figure*}
   \includegraphics[width=18cm, angle=-90]{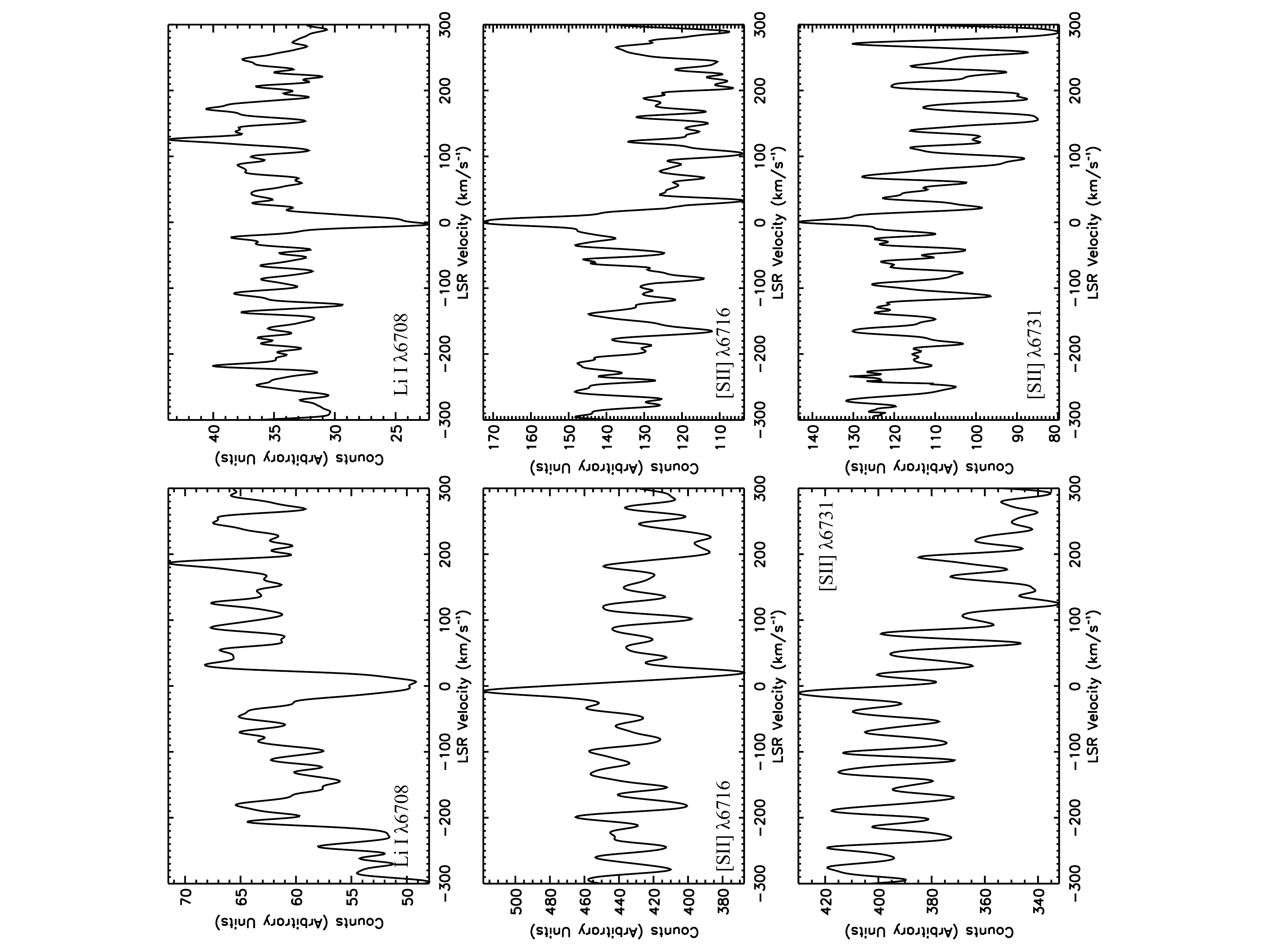}
     \caption{The Li I absorption line and the [SII]$\lambda\lambda$6716, 6731 doublet in the 2006 (left column) and 2008 (right column) spectra. The Li I spectrum is extracted from the source position and used to estimate the systemic velocity at $\sim$ +2~kms$^{-1}$. The [SII] lines are not detected at the source position but only by summing over $\pm$ 1 \arcsec. As explained in the text this is likely due to the lower critical density of the [SII] lines compared to the [OI]$\lambda$6300 line.}
     \label{image_comp}
\end{figure*}

\begin{figure*}
   \includegraphics[width=18cm]{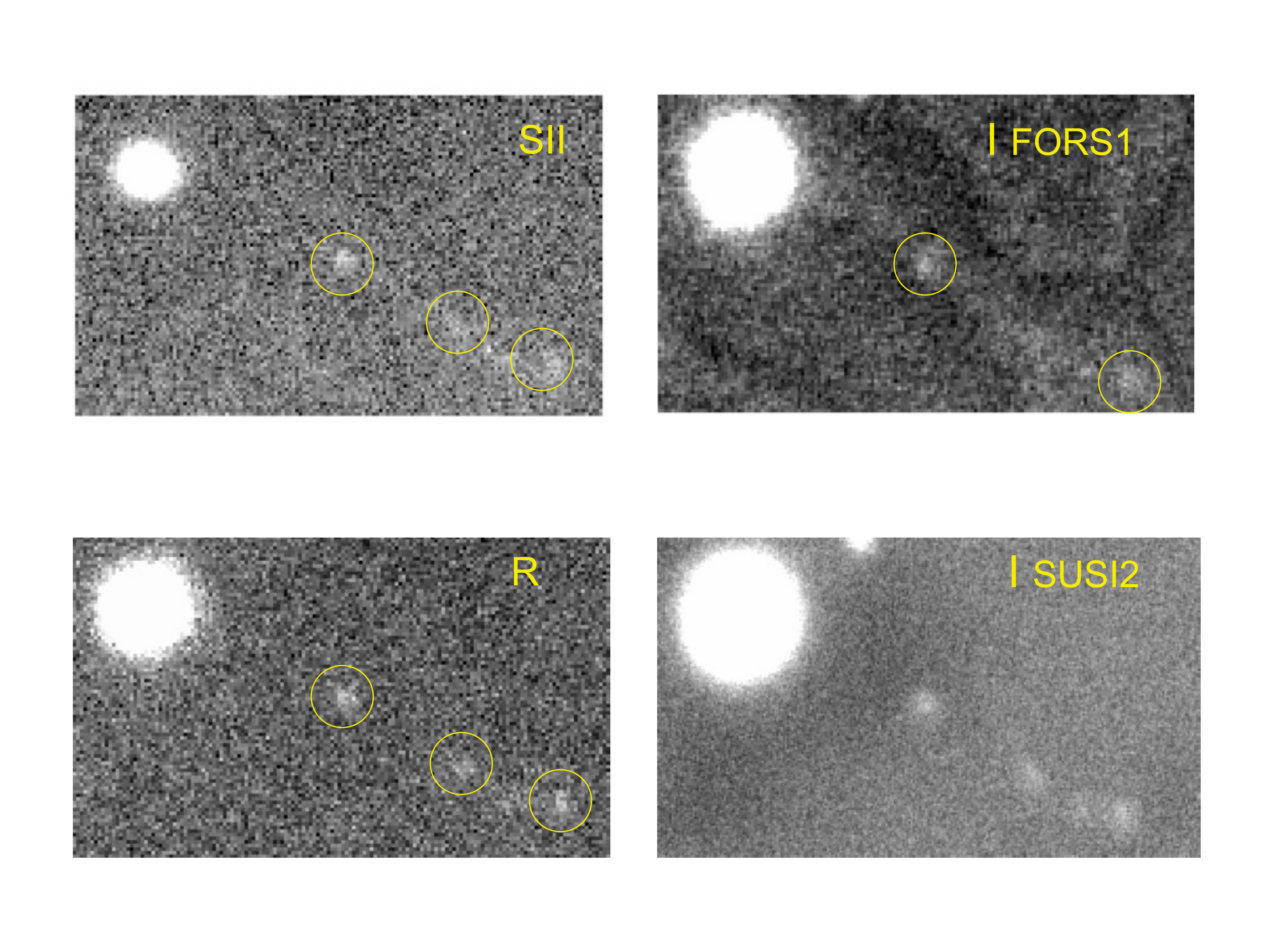}
     \caption{Here the knot-like features seen to the S-W of 2M1207A with a PA of $\sim$ 245$^{\circ}$ are compared in the [SII], R-band and I-band images. Two of the emission features are detected in the FORS1 I-band image. Also shown in a SUSI2 image taken two years previous to the FORS1 observations}
     \label{extended}
\end{figure*}

\begin{figure*}
   \includegraphics[width=18cm]{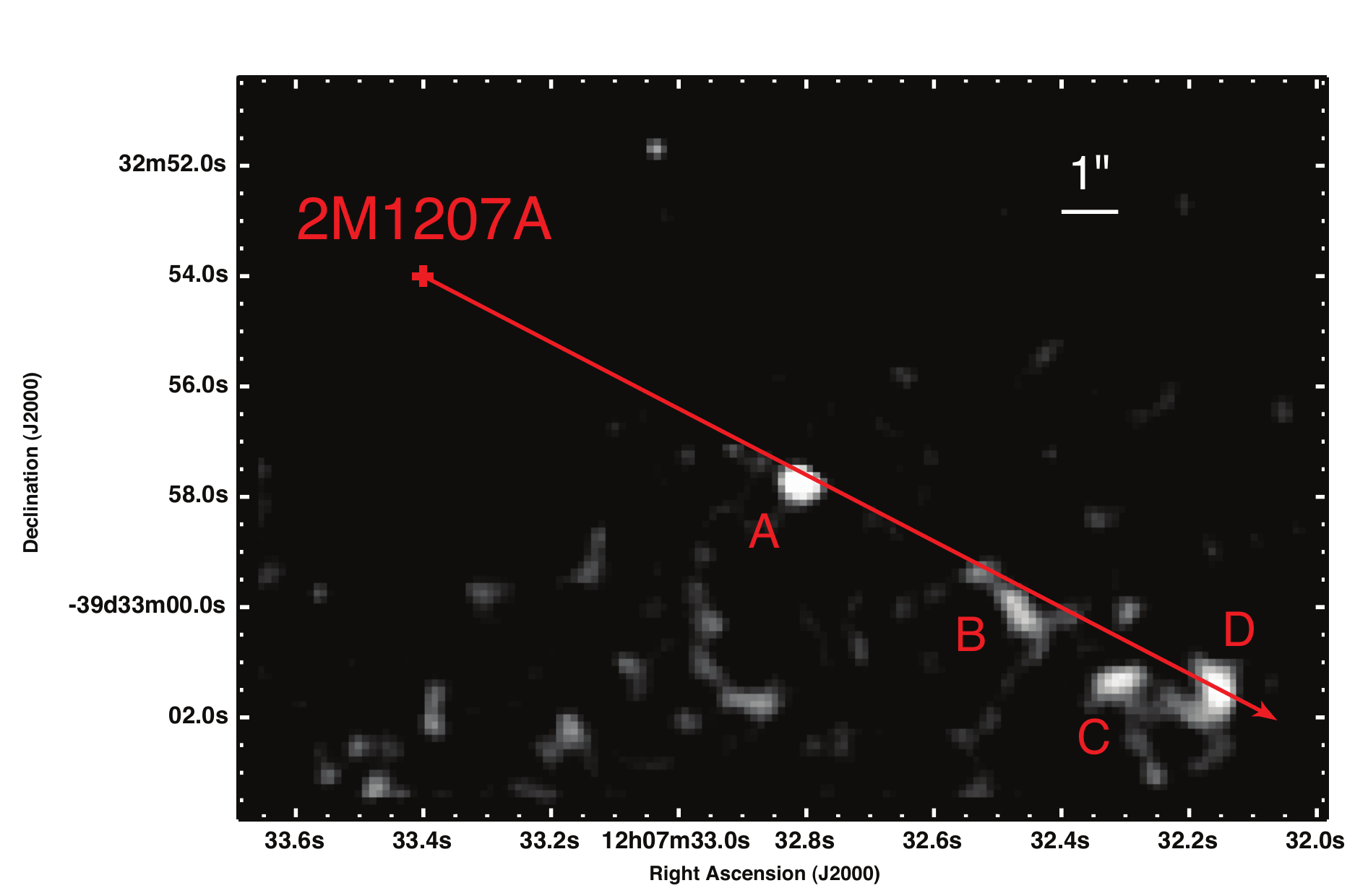}
     \caption{Gaussian smoothed [SII]-I image. 4 separate emission features are revealed with the final one having a clear bow-shock shape. The positions of the features are well fitted with a PA of 245$^{\circ}$ which is the same PA as estimated from the spectro-astrometry. }
     \label{smoothed}
\end{figure*}

\begin{figure*}
   \includegraphics[width=18cm]{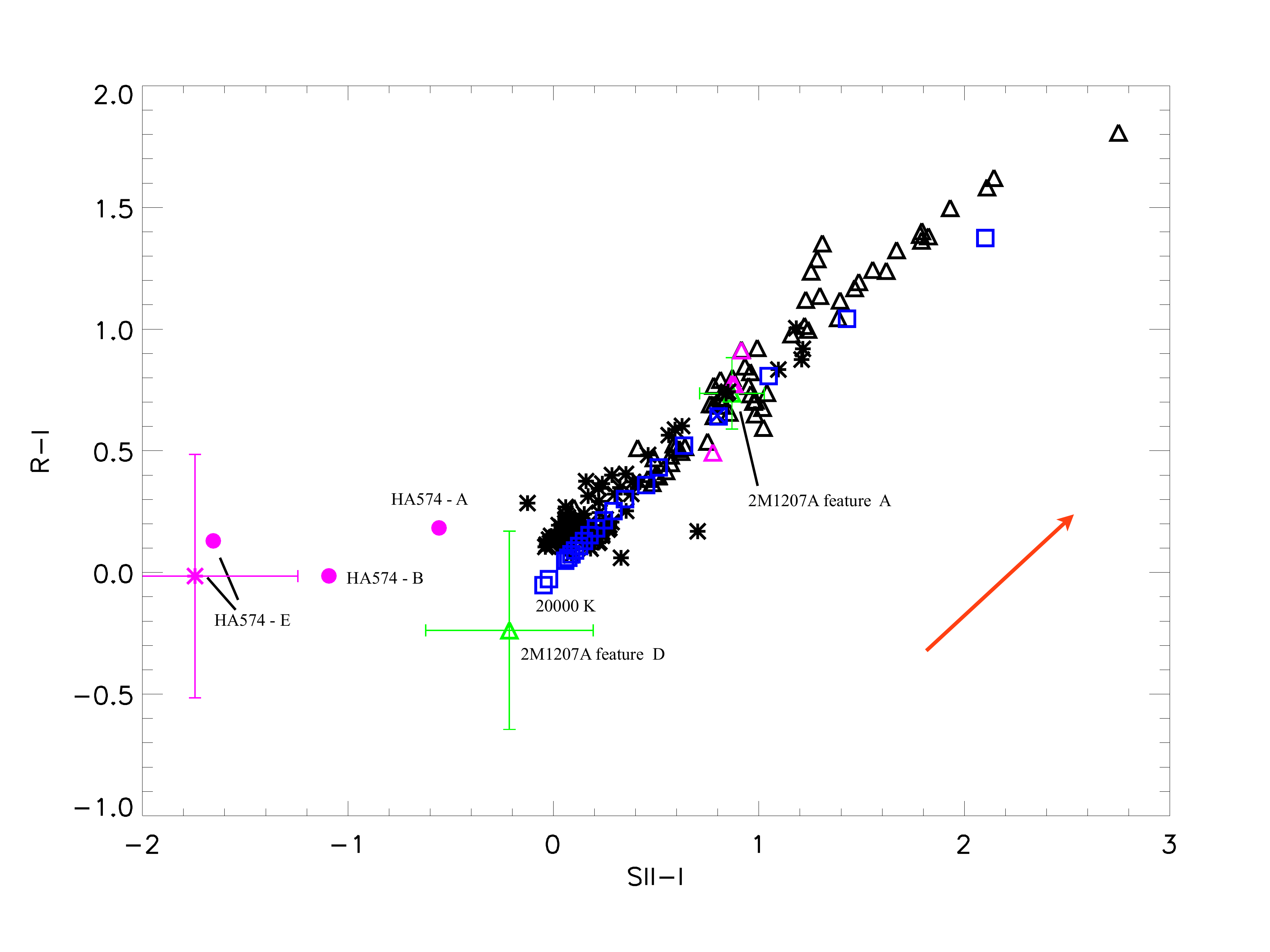}
     \caption{Using a color-color analysis to separate HH objects from stars and galaxies. The aim of this analysis is to compare the colors of the objects in the TW Hya images with synthetic colors and with an analysis of the protostar ESO-HA 574 and its associated HH objects, in order to test if the faint knot-like features {\it A} and {\it D} are also HH objects. The squares represent synthetic colors, the circles colors estimated from spectra, the triangles represent objects in TW Hya included in the FOV of the 2M1207 images and the asterisks represent objects in Cha I included in the FOV of the ESO-HA 574 images. The red arrow is the reddening vector. 
   Specifically the blue squares are synthetic colors for blackbodies in the range 20000~K to 1500~K, calculated in the range of the FORS1 filters. 
   The purple circles are the color estimates for the knots HA574-A, HA574-B and HA574-E, calculated from XSHOOTER spectra. The black asterisks are stars in Cha I and the purple asterisk marks the color of HA574-E measured from the image. The black triangles are stars in the TW Hya region, the purple triangles are identified galaxies and the green triangles features {\it A} and {\it D}.  The marked errors in the colors for HA574-E and {\it A}, {\it D} are quite large while the errors for the other points are found to be $<$ 0.02.  The position of feature {\it D}, the bow-shock shaped feature lies with the other known HH objects while {\it A} lies amongst the stars and galaxies. Based on this analysis and other evidence discussed we conclude that {\it D} is a shock in the 2M1207A outflow. }
     \label{color}
\end{figure*}

\end{document}